\documentclass[aps,prc,twocolumn,groupedaddress,showpacs,fleqn]{revtex4}
\usepackage{graphicx}
\usepackage{bm}
\usepackage{mathrsfs}
\newif\ifHIDEHIGHLIGNT
\HIDEHIGHLIGNTtrue %%%%% highlight the alterations
\ifHIDEHIGHLIGNT
\usepackage{ulem,color}
\newcounter{nnn}

\else

\fi
\begin{document}
\title{Farside-dominant quasinuclear rainbow 
in refractive $\alpha$+$\alpha$ scattering 
}

\author{S. Ohkubo}
\affiliation{Research Center for Nuclear Physics, Osaka University, 
Ibaraki, Osaka 567-0047, Japan }

\date{\today}

\pagenumbering{arabic}
\begin{abstract}
$\alpha$+$\alpha$ scattering has a long history since the first experiment by Rutherford and Chadwick in 1927 and has been studied thoroughly experimentally and theoretically. 
However, $\alpha$+$\alpha$ scattering has never been paid attention from the viewpoint of refractive scattering.
I have successfully analyzed the experimental angular distributions in $\alpha$+$\alpha$ scattering systematically over a wide range of incident energies $E_L$=53.4 - 280 MeV using a phenomenological optical model with a deep real potential. 
The existence of a farside-dominant quasinuclear rainbow with no well-defined rainbow angle and no supernumerary bow in the lit side followed by the shadow, which is not a genuine rainbow but a refractive scattering from a marginally small droplet at high energies,
is found for the first time in $\alpha$+$\alpha$ scattering. 
The refraction  due to  the deep potentials with an attractive core at short distances are discussed from the viewpoint of the Luneburg. 
The deep vs shallow problem of the potential and the nuclear rainbow  scattering in inelastic channels  are also discussed.
\end{abstract} 

\maketitle

\section{INTRODUCTION}
\par 
The rainbow on Earth has been attracting human attention for more than two thousand years, since long before the birth of modern science \cite{Newton,Airy1938,Nussenzveig1977,Adam2002,Maitte2005}. The mechanism of the rainbow due to refraction and reflection in a droplet was discovered by Descartes and then the origin of the beautiful colors was revealed by Newton \cite{Newton}. 
Airy \cite{Airy1938} found that the supernumerary bow, the Airy structure,
is due to the wave nature of light. Now the meteorological rainbow has been thoroughly understood as Mie scattering of electromagnetic waves from a droplet \cite{Nussenzveig1977,Adam2002}. Since the birth of quantum physics, it has been known that a rainbow is not only a macroscopic phenomenon in classical physics but also a microscopic phenomenon that occurs in the atomic and subatomic world \cite{Ford1959}. In fact, atomic rainbows \cite{Hundhausen1965} and nuclear rainbows \cite{Goldberg1972,Goldberg1974} have been  observed in experiments. The nuclear rainbow \cite{Khoa2007} that occurs only by refraction is a Newton zero-order rainbow, which Newton expected to exist in nature \cite{Newton}. Furthermore, in addition to  Newton zero-order primary nuclear rainbow, the existence of a secondary bow generated dynamically by a coupling to the inelastic channels has been reported \cite{Ohkubo2014,Ohkubo2015}.
The concept of a  rainbow, which is a wave phenomenon that appears on disparate scales  in physics, is not limited to the electromagnetic force and the nuclear force.
Recently, after the observation of gravitational waves \cite{Abbott2016}, rainbow scattering of gravitational waves from a compact body such as a neutron star or a black hole was investigated \cite{Dolan2017,Stratton2019}.

\par
Nuclear rainbows, which are observed under weak or incomplete absorption, have been extensively investigated experimentally and theoretically, especially for  systems including doubly closed magic nuclei, such as $\alpha$+$^{16}$O \cite{Michel1983}, $\alpha$+$^{40}$Ca \cite{Delbar1978},  and $^{16}$O+$^{16}$O \cite{Stiliaris1989,Nicoli1999,Khoa2000}, and also other systems including $\alpha$+$^{12}$C \cite{Ohkubo2002A}, $\alpha$+$^{90}$Zr \cite{Put1977,Ohkubo1995,Michel2000}, $^{16}$O+$^{12}$C \cite{Ogloblin1998,Nicoli2000,Szilner2001,Ogloblin2000,Ogloblin2003}, and $^{12}$C+$^{12}$C \cite{Stokstad1976,Bohlen1982,Bohlen1985,Michel2004}. The penetration of the projectile deep into the internal region of the target nuclei made it possible to determine the intercluster potential well up to the internal region. This made it possible to study the cluster structure of the compound system at  low excitation energies using the obtained intercluster potential, as shown typically for the $\alpha$+$^{16}$O \cite{Ohkubo1977,Michel1983} and the $\alpha$+$^{40}$Ca systems \cite{Michel1986,Michel1988,Ohkubo1988,Michel1998,Ohkubo1999}. 

A meteorological rainbow is smeared for a marginally small water droplet since the 
Airy structure in the lit side is obscured in Mie scattering. 
On the other hand, it has been unclear 
whether the nuclear rainbow persists or disappears for a small droplet  as in 
$\alpha$+$\alpha$ scattering.
Although $\alpha$+$\alpha$ scattering has been studied most thoroughly experimentally and theoretically \cite{Wheeler1941,Conzett1960,Igo1960,Darriulat1965,Bacher1972,%
Shimodaya1961,Shimodaya1962,Tamagaki1965,Saito1968,Tamagaki1968,Hiura1972,Tanabe1975,
Ali1966,%
Buck1977,Frisbee1972,Nadasen1978,Warner1994,Steyn1996,Rao2000} since the first experiment by Rutherford and Chadwick in 1927 \cite{Rutherford1927},
to the author's best knowledge, no attention has been paid from the viewpoint of refractive or rainbow scattering.

In the low energy region the phase shifts of $\alpha$+$\alpha$ scattering \cite{Conzett1960,Igo1960,Darriulat1965,Bacher1972} are described well 
microscopically in the resonating group method (RGM) \cite{Shimodaya1961,Shimodaya1962,Tamagaki1965,Tamagaki1968,Hiura1972}, semi-microscopically in the orthogonality condition model (OCM) \cite{Saito1968,Hiura1972,Tanabe1975} and phenomenologically 
with a potential model, a shallow potential with a repulsive core at short distances \cite{Ali1966} and a deep local potential \cite{Buck1977}.
The shallow potential with a repulsive core was mathematically shown to be a supersymmetric partner \cite{Baye1987} of the deep potential, in which the Pauli-forbidden states are embedded.
At high energies the angular distributions in $\alpha$+$\alpha$ scattering have been measured in the energy range 25 - 75 MeV/nucleon \cite{Darriulat1965,Frisbee1972,Nadasen1978,Warner1994,Steyn1996,Rao2000}.

\par

In this paper emergence of a farside-dominant quasinuclear rainbow, which is not a genuine nuclear  rainbow but a refractive scattering at high energies from a marginally small droplet without a well-defined rainbow angle,
% that {may} evolve into a nuclear rainbow for a larger droplet such as $^{12}$C, $^{16}$O and $^{40}$Ca,
is shown for the first time in $\alpha$+$\alpha$ scattering. Unlike  the genuine nuclear rainbow the farside-dominant quasinuclear rainbow is not accompanied by a supernumerary bow in the lit side followed by the shadow.

The paper is organized as follows. In Sec. II, after the characteristic feature of the experimental angular distributions in $\alpha$+$\alpha$ scattering at higher energies and the previous attempts to reproduce them are briefly reviewed, analysis of the experimental data using a two-range real potential allowing a different shape for the outer and inner regions is given.
Section III is devoted to the mechanism of the falloff and quasinuclear rainbows in refractive scattering.
In Sec. IV discussions on the obtained potentials from the viewpoint of the Luneburg lens, the deep vs shallow problem of the potential, bosonic effects, and nuclear rainbows in inelastic channels are presented. A summary is given in Sec. V.

\section{ANALYSIS OF $\alpha$+$\alpha$ SCATTERING}

%fig1 alpha+alpha falloff all data
\begin{figure}[t] 
\begin{center}
\includegraphics[keepaspectratio,width=8.6cm] {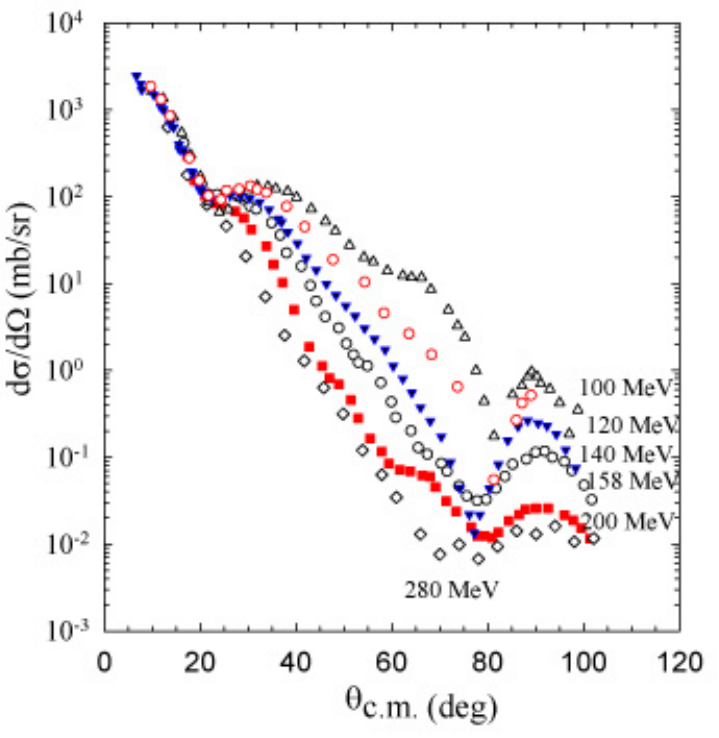}
\protect\caption{ {%(Color online) 
Experimental angular distributions in $\alpha$+$\alpha$ scattering at $E_L$=100, 120 \cite{Darriulat1965}, 140 \cite{Frisbee1972}, 158.2 \cite{Steyn1996}, 200 \cite{Steyn1996} and 280 MeV \cite{Rao2000} where the characteristic falloff of the cross sections is seen.
}}
\label{fig1}
\end{center}
\end{figure}

In Fig.~\ref{fig1} the experimental angular distributions in $\alpha$+$\alpha$ scattering at $E_L$=100 - 280 MeV are shown. The characteristic feature is that in all the angular distributions the cross sections rapidly decrease at around $\theta_{c.m.}$=20$^\circ$ toward large angles. 
At 120 MeV the cross sections in the falloff decrease almost four order of magnitude. Although the minimum at around 20$^\circ$ smears as the incident energy increases, the falloff evolves up to 280 MeV with a steeper slope.
In Ref.~\cite{Chauhan2009} $\alpha$+$\alpha$ scattering at $E_L$=100-280 MeV was studied using Glauber theory.
Since there is no concept of $\alpha$-$\alpha$ potential, the physical origin of the falloff is unclear.
The concept of $\alpha$-$\alpha$ potential, which was established to reproduce  the phase shifts in $\alpha$+$\alpha$ scattering below $E_L$=100 MeV \cite{Buck1977,Hiura1972,Tanabe1975}, seems useful also at  higher energies. 
Nadasen {\it et al.} \cite{Nadasen1978} tried to fit their measured angular distribution at 158.2 MeV by using a standard six-parameter Woods-Saxon optical potential as well as a nine-parameter Woods-Saxon optical potential with a two-range real attraction. They found that a better agreement with the experimental data was obtained 
with the latter potential with more parameters.
Warner {\it et al.} \cite{Warner1994} and Steyn {\it et al.} \cite{Steyn1996} analyzed 
their measured $\alpha$+$\alpha$ scattering angular distribution at 118  
and at 200 MeV, respectively.
Rao {\it et al.} \cite{Rao2000} analyzed their measured angular distribution at 280 MeV.
Farid \cite{Farid2006} studied these angular distributions by using a phenomenological potential and a folding model.
In all these analyses efforts were focused to reproduce the measured angular distributions phenomenologically and no attention was paid to the physical meaning of the observed characteristic angular distributions.

One notes that  the characteristic angular distributions with the falloff 
evoke nuclear rainbow scattering \cite{Khoa2007} since the systematic falloff of the cross sections occurs at  higher energies in the angular region where refractive scattering is classically forbidden. 

I analyze the observed angular distributions over a wide range of incident energies $E_L$=53.4 -280 MeV systematically from the viewpoint of $\alpha$-$\alpha$ potential.
I try to fit the experimental data in the frame of the optical potential model by taking a phenomenological approach.

 First my efforts were devoted to fitting the experimental angular distribution by using a Woods-Saxon squared potential, which had been very successful in describing the experimental angular distributions over a wide range of incident energies in $\alpha$ scattering from $^{16}$O \cite{Michel1983}, $^{40}$Ca \cite{Delbar1978} and $^{48}$Ca \cite{Ohkubo2020}. However, the angular distributions, especially at large angles, were not reproduced well.
A difficulty using a Woods-Saxon squared potential can be seen in Ref.~\cite{Farid2006}, in which an artificial multiplication, ranging 1.5 - 0.65, of the calculated cross sections was done without any justification to reproduce the experimental data individually at each energy. The Gaussian deep potential of Ref.~\cite{Buck1977} based on the RGM, which was successful at low energies \cite{Tan2013}, was also found to be unsuccessful at higher energies. Reference~\cite{Avrigeanu2003} reported that double folding model calculations using a density dependent effective two-body force such as DDM3Y, which had been very successful in $\alpha$ scattering from $^{16}$O \cite{Abele1993} and $^{40}$Ca \cite{Atzrott1996}, were unsuccessful in reproducing the angular distributions in $\alpha$+$\alpha$ scattering at lower energies $E_L$=8.87 - 29.5 MeV. 
Reference~\cite{Farid2006} adopted the effective two-body forces, which had not been extensively checked in typical $\alpha$ scattering such as $\alpha$+$^{16}$O and $\alpha$+$^{40}$Ca.

I take more
 a  flexible potential model approach to give a better fit than the conventional optical potential models
by allowing a different shape for the inner and outer regions of potential,
\begin {eqnarray}
&V(r) = V_1(r) + V_2 (r),
\label{Eq:pot}
\end {eqnarray} 
\noindent where 
$V_1(r) $ and $V_2 (r)$ represent the potentials in the outer and inner regions, respectively. I take $V_1(r) = - V_1 f(r; R_1, a_1)$ and $ V_2(r) = - V_2 f(r; R_2, a_2)$ with $f(r; R_i, a_i)$ being a Woods-Saxon form factor, which was first used in Ref. \cite{Nadasen1978}. The imaginary potential is assumed to be $W(r)=-W_0 f(r; R_W, a_W)$. 
The reduced radius $r_i$ is defined by $R_i=r_i 4^{1/3}$.
The Coulomb potential $V_{coul}(r)$ is assumed to be a uniformly charged sphere with $r_{c}$=1.3 fm. 
The two-range $\alpha$-$\alpha$ potential, which is $L$ (angular momentum) dependent, with an attraction at the outer region and a {\it} repulsion at short distances was used by Ali and Bodmer \cite{Ali1966} to reproduce the phase shifts for $L \le 4$ in $\alpha$+$\alpha$ scattering in the low energy region $E_L\leq 24$ MeV.
Equation~(\ref{Eq:pot}), which is $L$ independent, is an extension of the single-range deep attractive potential by Buck {\it et al.} \cite{Buck1977} to two ranges by (1) allowing the shape evolution of the potential at higher energies 
and (2) taking into account the effect of the Pauli principle at short distances $r\le$2 fm with $V_2(r)$.

%fig angular distributions fit 2 range WS1-Ws1Ws1 pot fit 
% 53.4, 77.55, 100, 120, 140, 158, 200, 280 MeV 
\begin{figure}[t]
\begin{center}
\includegraphics[keepaspectratio,width=8.6cm] {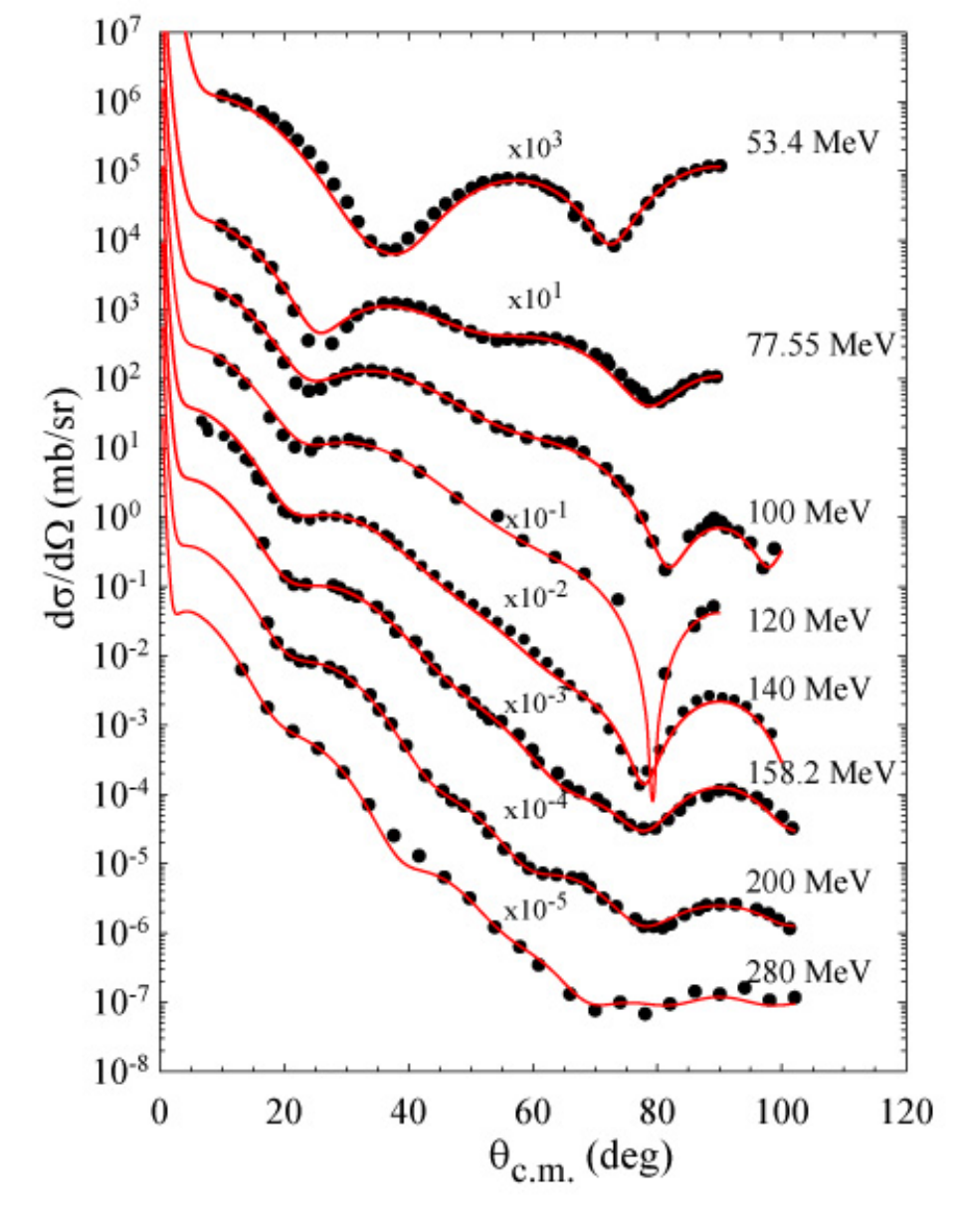}
\protect\caption{{
The angular distributions in $\alpha$+$\alpha$ scattering calculated with the
optical potentials
in Table~I (solid lines) are compared with the experimental data (closed circles) at $E_L$=53.4, 77.55, 100, 120 \cite{Darriulat1965}, 140 \cite{Frisbee1972}, 158.2, 200 \cite{Steyn1996} and 280 MeV \cite{Rao2000}.
}}
\label{fig:Ws1Ws1}
\end{center}
\end{figure}

%%Table I half column Ws1 WS1 WS1 53MeV - 280 MeV potential parameters 
\begin{table}[t]
\begin{center}
\protect\caption{ The optical potential parameters with a two-range Woods-Saxon form factor in $\alpha$+$\alpha$ scattering used in Fig.~\ref{fig:Ws1Ws1} and the volume integrals per nucleon pair $J_V$ for the real part of the potential.
$r_2$=0.545 except 0.531 at 200 MeV and 0.565 at 280 MeV, and 
$a_2$=0.142 except 0.152 at 200 MeV and 0.004 at 280 MeV.
Energies are in MeV, radii in fm and volume integrals in MeVfm$^3$.
}
\begin{tabular}{lccclcrcc}
\hline
\hline 
$E_L$ & $V_1$ & $r_1$ & $a_1$ & $V_2$ & $J_V$ & $W_0$ & $r_W$ & $a_W$ 
\\
\hline
53.4& 68.6& 1.622& 0.616&66.27& 494 &4.39 &2.149 &0.259 
\\ 
77.55 & 60.0 & 1.628 & 0.613 &55.0&435 &6.00 &2.149 &0.467 
\\ 
100 & 55.94 & 1.628 &0.613 & 66.53&408 & 6.71 &2.094 &0.467 % My pot
\\ 
120 & 52.94 & 1.628& 0.613 & 57.67 &385 & 6.97 &2.094 &0.467 % My pot
\\ 
% 140 MeV set 6
140 & 53.38 & 1.628&0.613 & 43.92& 384 & 7.48 &2.188 &0.618 
\\
158.2 & 53.75 & 1.628& 0.613 & 43.97& 387 & 9.62&2.094 & 0.467 
\\ % Steyn pot
200 & 53.36 & 1.595& 0.585 & 30.46&353 & 8.71&2.120 &0.410 
\\ % Steyn pot
280 & 53.00 & 1.595& 0.585 & 14.05&347 &9.20 &1.930 & 0.337 
\\ 
\hline
\hline
\end{tabular}
\end{center}
\label{Table:Ws1Ws1}
\end{table}

The experimental angular distributions are analyzed starting from the potential at 158.2 MeV in Ref.~\cite{Nadasen1978}. 
In Fig.~\ref{fig:Ws1Ws1} the calculated angular distributions are compared with the experimental data at $E_L$=53.4 - 
280 MeV. The optical potential parameters used in the calculations are listed Table~I. 
The calculations reproduce the experimental data well over a wide range of incident energies. One sees that the falloff of the experimental angular distributions above $E_L$=100 MeV are reproduced well. The peaks and valleys at large angles
beyond the falloff are also reproduced well. The experimental angular distributions at 53.4 and 77.55 MeV, where diffractive scattering dominates and had been used in the phase shift analyses in Ref.~\cite{Darriulat1965}, are reproduced well by the calculations. 

\section{THE FALLOFF AND REFRACTIVE SCATTERING}
\par
\begin{figure}[t!]
\begin{center}
\includegraphics[keepaspectratio,width=8.6cm]{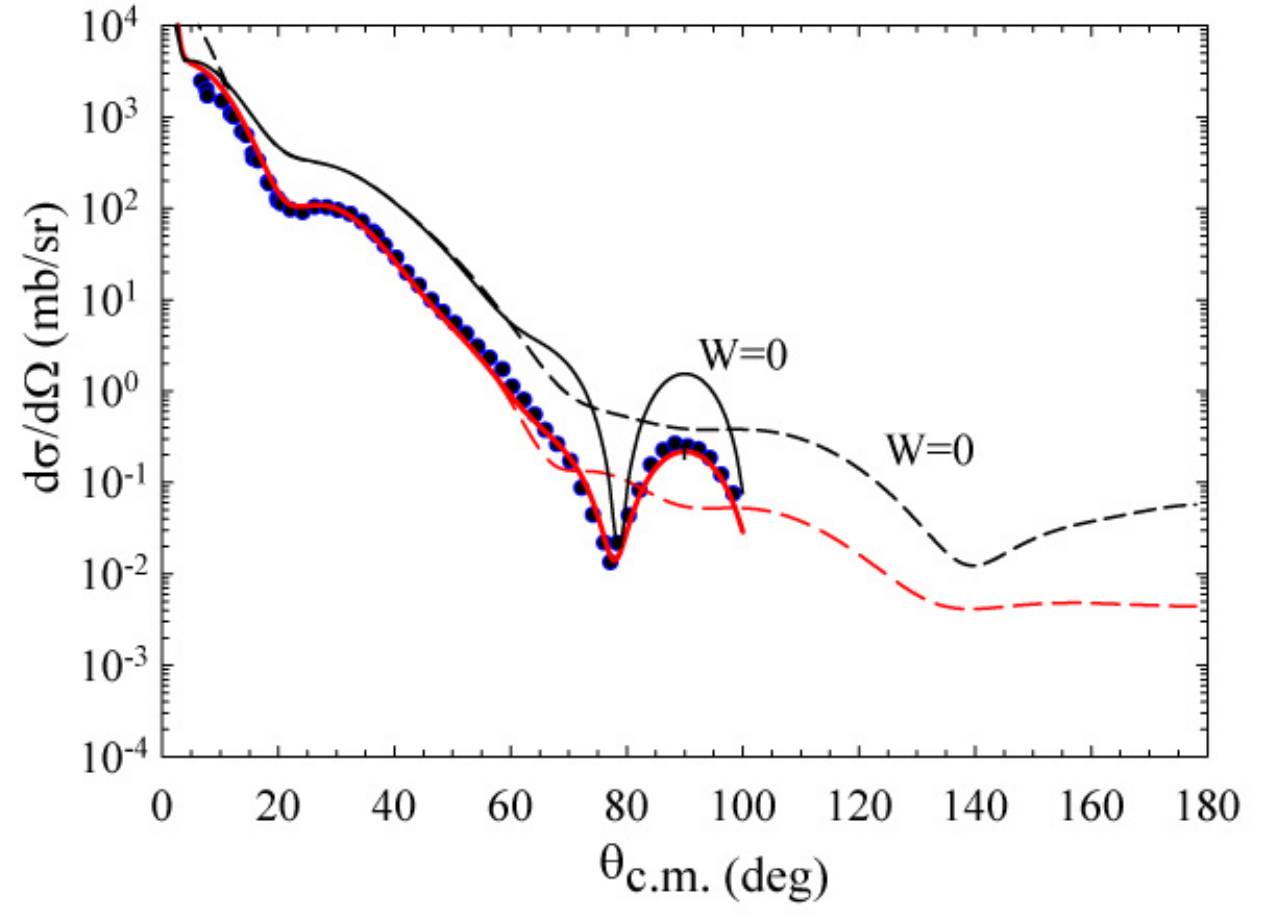}
\protect\caption{\label{fig:E=140MeVW=0} {
Comparison of the angular distributions in $\alpha$+$\alpha$ scattering at $E_L$=140 MeV calculated with (red solid line) and without (black solid line) the imaginary potential in Table I. The calculated unsymmetrized angular distributions with and without the imaginary potential are also displayed by the dashed lines together with the experimental data (closed circles) \cite{Frisbee1972}.
}}
\end{center}
\end{figure}

In order to see that the falloff of the angular distributions is due to the real part of the potential and not due to absorption, in Fig.~\ref{fig:E=140MeVW=0} the angular distributions calculated by switching off the imaginary potential are displayed at 140 MeV in comparison with the result with absorption in Fig.~2. One sees that the shapes of the two angular distributions with and without the absorption are similar,  and the falloff does not disappear by switching off the imaginary potential. This shows that the falloff pattern is not due to absorption. This is also the case for other energies $E_L\ge$100 MeV. Thus it is found that the falloff is due to the real part of the optical potential.
It is noted that although the two curves calculated with and without symmetrization differ drastically at large angles around 90$^\circ$, they differ little at $\theta_{c.m.}\le 60^\circ$.
This shows that the falloff pattern at $\theta\le 60^\circ$ is scarcely affected by boson symmetrization of the two $\alpha$ clusters.

%%fig3. 140MeV W=0 comparison fit Steyn type pot searched 
% to show that the falloff is due to the real potential and not due to absorption

\begin{figure}[t!]
\begin{center}
\includegraphics[keepaspectratio,width=8.6cm]{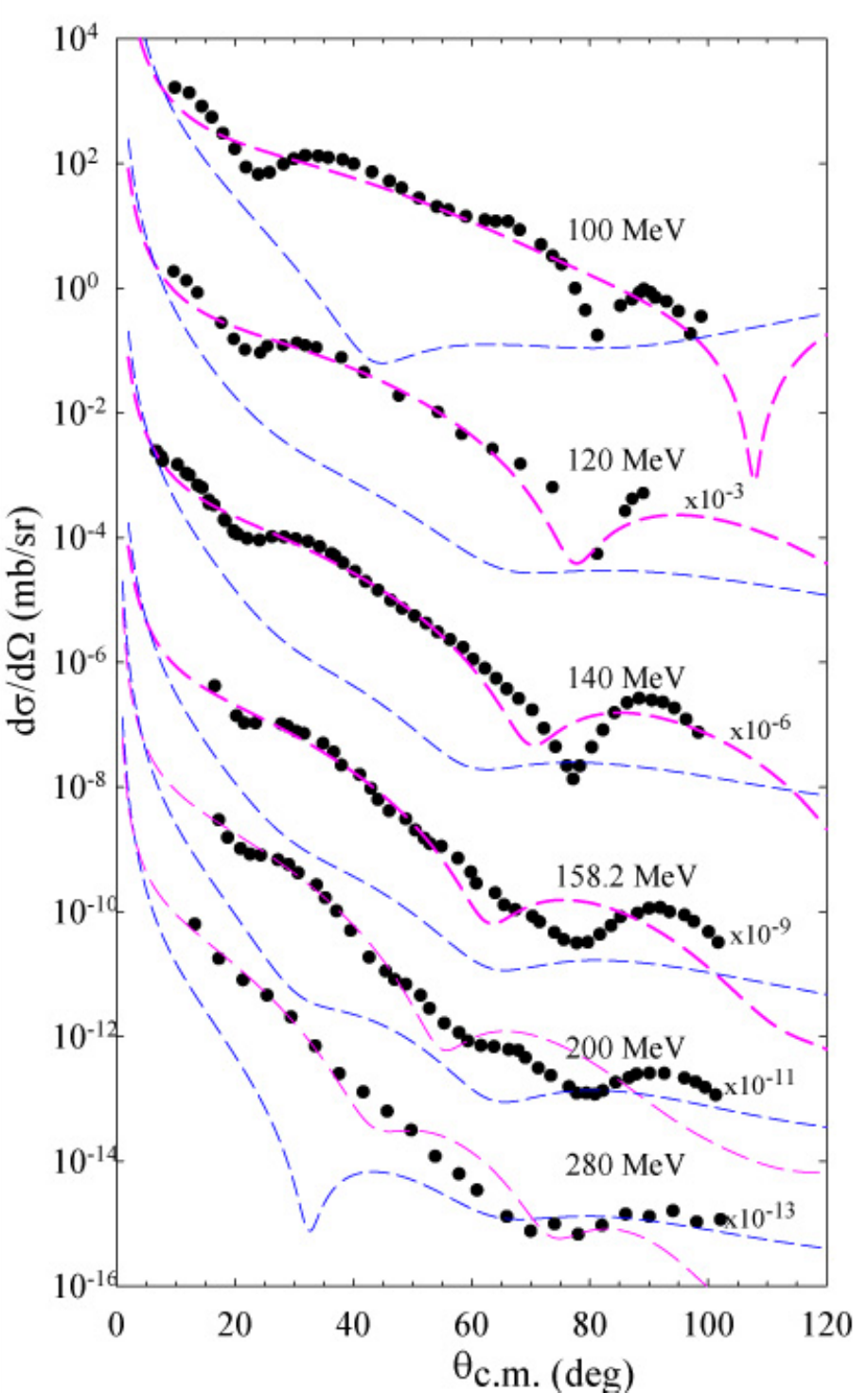}%no Total cal
\protect\caption{ { The calculated angular distributions in $\alpha$+$\alpha$ scattering at $E_L$=100 - 280 MeV in Fig.~\ref{fig:Ws1Ws1} 
are decomposed into the farside (pink long dashed line) and nearside (blue medium dashed line) components, which are not symmetrized. 
The experimental data (closed circles) are shown for comparison.
}}
\label{fig:NFall}
\end{center}
\end{figure}

To understand clearly the physical meaning of the falloff, in Fig.~\ref{fig:NFall} 
the angular distributions calculated with the potentials in Table~I are decomposed into the farside and nearside components using the prescription in Ref.~\cite{Fuller1975}. 
The farside and nearside decomposition has been very powerful \cite{Khoa2007}  for  understanding the underlying mechanism of the characteristic features of the angular distributions in  scattering involving not only nonsymmetric systems such as $\alpha$+$^{16}$O \cite{Hirabayashi2013}, $^{16}$O+$^{12}$C \cite{Ohkubo2014}, and $^{13}$C+$^{12}$C \cite{Ohkubo2015} but also symmetric two-bosonic systems such as $^{12}$C+$^{12}$C \cite{McVoy1992} and $^{16}$O+$^{16}$O \cite{Khoa2000,Michel2000A,Michel2001}.
One sees that the farside scattering dominates in the angular region of the falloff where boson symmetrization gives little effect, as noted in Fig.~3. 
On the other hand, in the falloff region the nearside contribution (blue medium dashed line) is much smaller than the farside contribution. Thus it is confirmed that the falloff is caused by the refractive farside scattering due the real potential. 
Also note that there appear no primary maximum (rainbow) and primary minimum in the angular region just before the falloff starts in the farside scattering. The primary maximum and minimum do not appear even in the calculations with $W=0$. This means that there are no supernumerary Airy maximum and minimum in the refractive $\alpha$+$\alpha$ scattering. If the falloff is due to a genuine nuclear rainbow, the primary Airy maximum, which is the bright side of the rainbow, should appear. Therefore this refractive scattering is not a genuine nuclear rainbow scattering. Although the falloff corresponds to the dark shadow classically forbidden, it is not the darkside of the genuine rainbow.

%%fig4 N/F 100-280MeV systematics N/F decomposition
\par
\begin{figure}[t!] 
\begin{center}
\includegraphics[keepaspectratio,width=8.6cm] {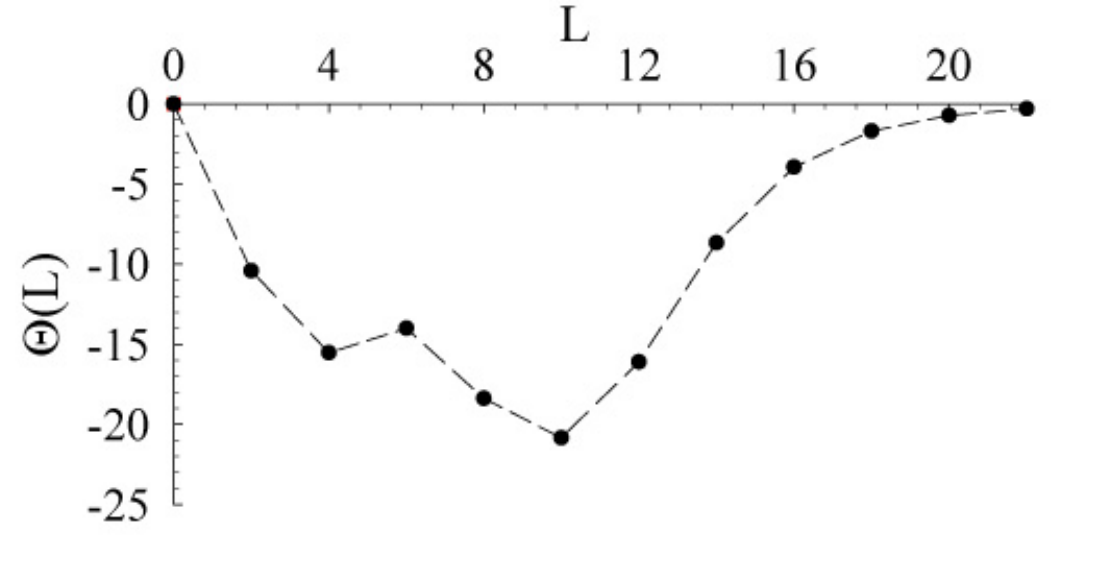}
\protect\caption{ {%(Color online) 
Illustrative deflection function without a rainbow angle. $\Theta$($L$)=2$d \delta_L/d L$ in $\alpha$+$\alpha$ scattering at 
280 MeV calculated for the the potentials in Table~I, where $\delta_L$ is the phase shift. The line is to guide the eye.
}}
\label{fig:deflection}
\end{center}
\end{figure}

The appearance of the lit side with no Airy structure and the dark side with the falloff in the refractive scattering may be understood if one looks into the deflection function $\Theta$($L$)=2$d \delta_L/d L$ in the classical picture where $\delta_L$ is the phase shift.
In Fig.~\ref{fig:deflection}  I show $\Theta$($L$) for the real part potential 
at $E_L$= 280 MeV in Table~I calculated approximately by the differences of $\delta_L$ with respect to $L$. 
At $E_L$=280 MeV, center-of-mass energy $E_{c.m.}$=140 MeV, the partial waves $L\le$30 are involved in refractive scattering. However, because of boson symmetrization the number of  partial waves physically involved is drastically reduced by  half, which makes the function $\Theta$($L$) sparse and less smooth.
One of the characteristic features of $\alpha$+$\alpha$ scattering is this Bose statistics.
The concept of a deflection function has physical meaning at the incident energies where the semiclassical picture works \cite{Ford1959}. In Fig.~\ref{fig:deflection}, there is no stationary minimum and no rainbow angle. 
Also around the minimum the number of  involved partial waves
is too small, only a few, to be approximated by a continuous parabola that creates the Airy structure in the lit side. 
On the other hand, the falloff, the dark side, appears since refractive scattering to the angular region beyond the minimum is classically forbidden as seen in Fig.~\ref{fig:deflection}, 
irrespective of the absent  rainbow angle. This is the reason why the falloff appears clearly in the experimental angular distributions over a wide range of incident energies $E_L\ge$100 MeV as seen in Fig.~\ref{fig1}. 

%Fig.5

Although no Airy structure appears in the calculated farside scattering, one sees a primary maximum at the angle just before the falloff in the experimental angular distributions in Fig.~\ref{fig1}. 
The primary maximum as well as the primary minimum arise as a consequence of the interference between the farside and nearside scatterings.
They move forward as the incident energy increases similarly to the Airy maximum and minimum $A1$ of the genuine nuclear rainbow. Thus it is found the refractive $\alpha$+$\alpha$ scattering at high energies is not a genuine nuclear rainbow but a farside-dominant quasinuclear rainbow scattering.

\section{DISCUSSIONS}

\begin{figure}[t]
\begin{center}
\includegraphics[keepaspectratio,width=7.6cm] {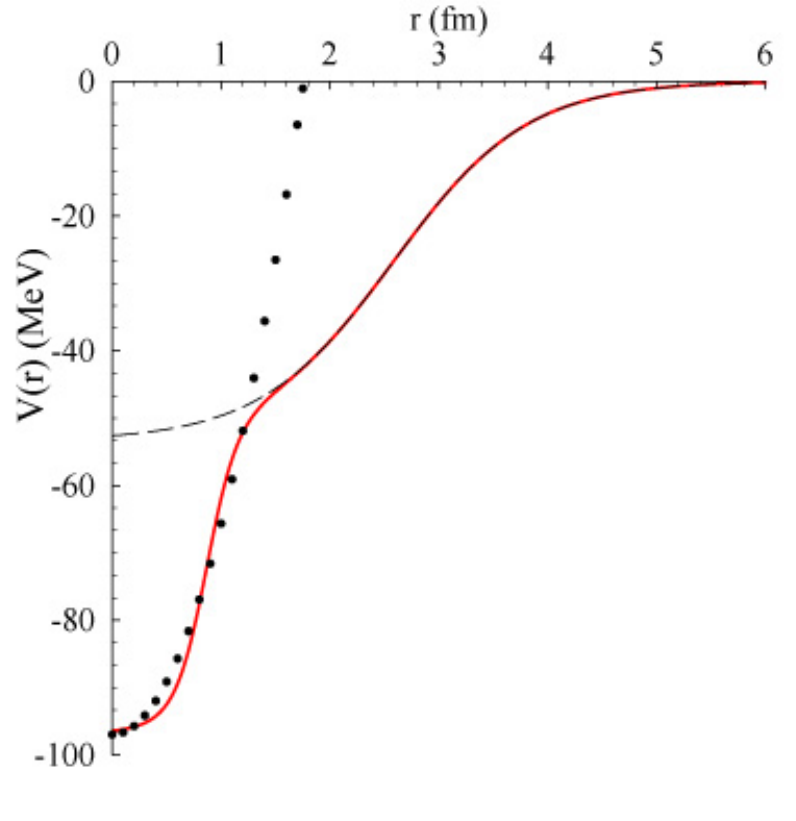}
\protect\caption{ {The potential $V(r)$ at 140 MeV (solid line) together with the long-range part $V_1(r)$ (dashed line)
in Eq.(\ref{Eq:pot}) are displayed. 
The Luneburg lens potential with $V_0$=97 MeV and $R_0$=1.76 fm is displayed by the dots.
}}
\label{fig:140MeV-V1V2}
\end{center}
\end{figure}

That $\alpha$+$\alpha$ scattering at high energies is different from the ordinary nuclear rainbow scattering can be also understood from the viewpoint of a nuclear lens. As discussed by the present author and his collaborators in Ref.~\cite{Michel2002}, the nuclear rainbow is an astigmatism phenomenon by the lens that resembles a Luneburg lens potential, a truncated harmonic oscillator potential given by $V(r)=V_0 ((r/R_0)^2-1)$ for $r \le R_0$ and $V(r)=0$ for $r>R_0$, where $R_0$ is the size of the lens and the focus is $R_f =R_0 \sqrt{E_{c.m.}/V_0}$.
In fact, the potentials for $\alpha$+$^{16}$O, $\alpha$+$^{40}$Ca, $^{16}$O+$^{16}$O, and $^{16}$O+$^{12}$C resemble the Luneburg lens potential except in the diffuse surface region \cite{Ohkubo2016,Michel2002}, which causes astigmatism, i.e., a  nuclear rainbow. 
On the other hand, the potentials for $\alpha$+$\alpha$ scattering at the higher energy region do not resemble a Luneburg lens. In Fig.~\ref{fig:140MeV-V1V2} the potential at 140 MeV is displayed. One sees that the Luneburg potential which simulates the internal region of the potential (displayed by the closed circles) is completely different from the potential $V(r)$. 
Although the potentials $V_1(r)$ and $V_2(r)$ are individually simulated approximately by a Luneburg lens potential except for the diffuse surface with $V^{(1)}_0$=45 MeV and $R^{(1)}_0$=1.21 fm with its focus $R^{(1)}_f$=0.97 fm, and $V^{(2)}_0$=53 MeV and $R^{(2)}_0$=3.75 fm with its focus $R^{(2)}_f$=3.26 fm, the total potential $V(r)$ does not behave like a Luneburg lens. This means that, unlike the ordinary nuclear rainbow refracted by a single Luneburg-lens-like potential, the incident projectile of  the 
farside-dominant quasinuclear rainbow is refracted by the two Luneburg-lens-like potentials with  different focuses.

As for the real potential, there has been a long-standing question whether the real potential is deep or shallow in $\alpha$ scattering and in heavy ion scattering. 
As mentioned in Sec. I, although for the most typical $\alpha$+$\alpha$ system a shallow potential accompanied with a repulsive core at short distances \cite{Ali1966}
has been widely used, extensive and systematic study of nuclear rainbow scattering \cite{Khoa2007} in systems such as $\alpha$+$^{16}$O, $\alpha$+$^{40}$Ca, $^{16}$O+$^{16}$O and other heavy ion systems by using a deep potential, which has a
 Luneburg-lens-like potential in the internal region \cite{Ohkubo2016}, showed that the deep potential is favored over the shallow potential despite the supersymmetric equivalence \cite{Baye1987} of the two potentials.
There may be an argument that, since a potential is not an observable in quantum mechanics, it is not meaningful to ask which is favoered, a deep attractive potential or a shallow potential accompanied with a repulsive core. 
However, in classical or semiclassical phenomena such as the nuclear rainbow scattering, a potential has a physical meaning, that is, an attractive potential (force) causes refraction and a shallow potential (force) accompanied with a repulsive core at short distances causes reflection at short distances.
The $L=0$ wave, which is not hampered by the centrifugal barrier, can penetrate deep into the very internal region where the repulsive potential dominates due to the Pauli principle in the shallow potential picture and can yield  information about the potential at $r<2$ fm.
In the present $\alpha$+$\alpha$ system, absorption is exceptionally weak even for the $L=0$ wave at the higher energies of refractive scattering; for example, the modulus of the $S$ matrix is  $S_L$=0.4 at $E_L$=140 MeV ($E_{c.m.}$=70 MeV) compared with $S_L=0.06-0.08$ for $\alpha$+$^{16}$O scattering at $E_L$=49.5 MeV ($E_{c.m.}$=39.6 MeV) and 69.5 MeV ($E_{c.m.}$=55.6 MeV) \cite{Michel1983}.
In the present analysis,
in order to reproduce the behavior of the experimental data precisely, not a repulsive potential but an additional deep potential $V_2(r)$ is needed at $r<2$ fm. This reinforces that a deep potential is favored.
To the author's best knowledge up to now it has not been reported that the  present refractive $\alpha$+$\alpha$ scattering can be systematically and precisely reproduced by a shallow potential accompanied with a repulsive core at short distances as in  the Ali-Bodmer potential \cite{Ali1966}.
Why the deep potentials for $\alpha$+$\alpha$ scattering at the higher energies are very different from the Luneburg lens is important and interesting, but is not a subject of the present paper and will be discussed elsewhere in a separate paper. 
\par

\par
I briefly discuss the origin of the deep minimum 
at 70-85$^\circ$ beyond the falloff in Fig.~\ref{fig1}. As displayed at 140 MeV in Fig.~\ref{fig:E=140MeVW=0}, the calculations with and without absorption both locate a minimum at around 80$^\circ$, in agreement with the experimental data. On the other hand, unsymmetrized calculations both with and without absorption do not show any deep minimum at around 80$^\circ$. Thus 
the deep minimum arises as a consequence of symmetrization due to the bosonic statistics of the two-$\alpha$ system. 
This symmetrization minimum at large angle beyond the falloff has been clearly observed because of the weak absorption for the $\alpha$+$\alpha$ system.
It is intriguing whether a similar symmetrization minimum beyond the falloff can be systematically  observed at high energies in $^{16}$O+$^{16}$O rainbow scattering.

%fig V1V2 140MeV pot Luneburg

When nuclear rainbows and the Airy structure are observed in elastic scattering, they are also observed in inelastic scattering under incomplete absorption for many systems, for example $\alpha$+$^{12}$C(0$_2^+$; 7.65 MeV) \cite{Ohkubo2002A}, $^3$He+$^{12}$C(0$_2^+$) \cite{Hamada2013}, $\alpha$+$^{16}$O(3$^-$; 6.13 MeV) \cite{Hirabayashi2013}, $\alpha$+$^{40}$Ca(3$^-$; 3.74 MeV) \cite{Michel2001B}, and 
$^{16}$O+$^{12}$C(2$^+$; 4.44 MeV) \cite{Ohkubo2014C}.
The Feshbach resonance state $^4$He($0_2^+$, 20.2 MeV) \cite{Horiuchi2008} at the highly excited energy near the $3N$+$N$ threshold has an extended cluster structure, i.e. an extended nuclear lens, similar to the Hoyle state $^{12}$C(0$_2^+$; 7.65 MeV).
In inelastic $\alpha$+$^4$He($0_2^+$) scattering odd parity partial waves, which are not allowed in elastic $\alpha$+$\alpha$ scattering because of bosonic statistic, are available.
It is highly desired to observe inelastic $\alpha$+$^4$He($0_2^+$) scattering in experiments to see whether a genuine nuclear rainbow exists in the inelastic channel.

\section{SUMMARY}
To summarize, I have discovered a farside-dominant quasinuclear rainbow without Airy structure for the first time 
in refractive $\alpha$+$\alpha$ scattering. This was done by analyzing $\alpha$+$\alpha$ scattering systematically over a wide range of incident energies $E_L$=53.4-280 MeV by using a phenomenological potential with  different shapes for the inner and the outer regions.
The experimental angular distributions are reproduced well by the calculations. 
The emergence of the farside-dominant quasinuclear rainbow in $\alpha$+$\alpha$ scattering is caused by the inter-nucleus potential, which differs from a Luneburg lens for a marginally small projectile and target under the quantum effect of bosonic statistics.
The present study in refractive $\alpha$+$\alpha$ scattering in the high energy region, which is sensitive to the very internal region $r\le2$ fm of the real potential, reinforces that a deep attractive potential is favored to explain the refractive phenomenon despite the suspersymmetric equivalence of the deep potential and the shallow potential accompanied with the repulsive core at short distances. I propose an experimental search of genuine nuclear rainbows with a supernumerary Airy structure in the lit side in inelastic $\alpha$+$\alpha$ scattering where odd parity partial waves can be involved because of no bosonic symmetrization.

\begin{acknowledgements}
The author thanks the Yukawa Institute for Theoretical Physics, Kyoto University for the hospitality during a stay in 2023. He also thanks 
%S. R. Dolan for interests in the Luneburg lens approach and 
Y. Hirabayashi for discussions. 
\end{acknowledgements}

\end{document}